\documentstyle[prl,preprint,aps]{revtex}
\begin{document}
\draft
\title{TWO-PION EXCHANGE NUCLEAR POTENTIAL - \\
CHIRAL CANCELLATIONS}
\author{J-L. Ballot}
\address{Division de Physique Th\'eorique,
Institut de Physique Nucl\'eaire\thanks{Unit\'e de Recherche des
Universit\'es Paris 6 et Paris 11 associ\'ee au CNRS.}\\
F91406 Orsay C\'edex, France, e-mail:ballot@frcpn11.in2p3.fr}
\author{M.R. Robilotta}
\address{Instituto de F\'{\i}sica, Universidade de S\~ao Paulo \\
C.P. 20516, 01452-990, S\~ao Paulo, SP, Brazil,
e-mail: robilotta@if.usp.br}
\author{C.A. da Rocha}
\address{Instituto de F\'{\i}sica Te\'orica,
Universidade Estadual Paulista \\
R. Pamplona, 145 - 01405-900 - S\~ao Paulo, SP, Brazil,
e-mail: carocha@vax.ift.unesp.br}
\date{\today}
\maketitle
\begin{abstract}
We show that chiral symmetry is responsible for large cancellations in the
two-pion exchange nucleon-nucleon interaction, which are similar to those
occuring in free pion-nucleon scattering.
\end{abstract}

The two-pion exchange nucleon-nucleon potential ($\pi\pi$E-NNP) in the
framework of chiral symmetry has recently attracted considerable attention,
especially as far as the restricted pion-nucleon sector is concerned
\cite{OVK92,CPS92,Bir94,FC94,RR94}.
This process is closely related to the pion-nucleon scattering amplitude, as
pointed out many years ago by Brown and Durso \cite{BD71}. In the non-linear
realization of chiral symmetry, one possibility is to write the pion-nucleon
interaction Lagrangian as a sum of scalar and pseudoscalar terms, as follows:
\begin{equation}
{\cal L}_{PS+S} = \cdots - g \overline{N}\,
  \left[\sqrt{f_{\pi}^2 - \phi^2} + i \bbox{\tau} \cdot \bbox{\phi}
  \gamma_5\right] N+\cdots \;,
\label{Eq.1}
\end{equation}
\noindent
where $g$ is the $\pi$N coupling constant, $f_\pi$ is the pion decay constant
and $N$ and $\phi$ are the nucleon and pion fields, respectively.

In the case of pion-nucleon scattering, this Lagrangian yields a tree amplitude
which contains poles in the $s$ and $u$ channels, as well as a scalar contact
interaction, which is the signature of chiral symmetry. At low energies, this
last term cancels a large part of the pole amplitude, giving rise to a final
result which is much smaller than the individual contributions.

As far as nucleon-nucleon scattering is concerned, the two-pion exchange
amplitude to order $g^4$ is given by five diagrams, usually named box,
crossed box, triangle (twice) and bubble \cite{RR94}.
The first two diagrams contain only
nucleon propagators and are independent of chiral symmetry, whereas the
triangles and the bubble involve the scalar interaction and hence are due to
the symmetry. When one considers the potential instead of the amplitude, the
iterated OPEP has to be subtracted from the box diagram.

In this work we show that, as in pion-nucleon scattering, there are large
cancellations among the various individual contributions to the interaction,
that yield a relatively small net result, and thus prevent the perturbative
explosion of the amplitude.

Using the potential in coordinate space produced
recently \cite{RR94} and parametrized in Ref.~\cite{RR95}, it is possible to
notice two important cancellations within the scalar-isoscalar sector of the
$\pi\pi$E-NNP. The first of them happens between the triangle and bubble
contributions, as shown in Fig.~\ref{Fig.1}. It is worth noting that the scale
of this figure is given in thousands of MeV. The other one occurs when the
remainder from the previous cancellation (S) is added to the sum of the box
and crossed box diagrams (PS). In this last case, the direct inspection of the
profile functions for the potential, given in Fig~\ref{Fig.2}, provides
just a rough estimate of the importance of the cancellation,
since the iterated OPEP is not included.
Therefore the second cancellation can be better studied
directly in the NN scattering problem, by considering the singlet waves, where
the influence of chiral symmetry is stronger.

It is well known that a full chiral calculation of the potential in momentum
space requires the inclusion of undetermined counterterms in the Lagrangian,
involving higher orders of the relevant momenta \cite{OVK92}.
In configuration space, on the
other hand, these counterterms become delta functions which affect only the
origin and hence are effective just for waves with low orbital momentum.
Therefore, in order to avoid these undetermined short range effects, we
consider the behaviour of the $^1D_2,\;^1G_4,\;^1F_3$, and $^1H_5$ waves.

For each channel, we decompose the full NN potential $V$ as
\begin{equation}
V=U_\pi+U_{PS}+U_S+U_C\;,
\label{Eq.2}
\end{equation}
\noindent
where $U_\pi$ is the OPEP, $U_C$ represents the short ranged core
contributions, $U_{PS}$ is due to the box and crossed box diagrams whereas
$U_S$ is associated with the chiral triangle and bubble interactions. Using
the variable phase method, it is possible to write the phase shift for
angular momentum $L$ as \cite{Cal63,BR94}
\begin{equation}
\delta_L=-\frac{m}{k}\int_0^\infty dr\;V\;P_L^2\;.
\end{equation}
\noindent
In this expression, the structure function $P_L$ is given by
\begin{equation}
P_L=\hat{\jmath}_L\,\cos{D_L} -\hat{n}_L\,\sin{D_L}\;,
\end{equation}
\noindent
where $\hat{\jmath}_L$ and $\hat{n}_L$ are the usual Bessel and Neumann
functions
multiplied by their arguments and $D_L$ is the variable phase. Using the
decomposition of the potential given in Eq.~\ref{Eq.2}, one writes the
perturbative result
\begin{eqnarray}
\delta_L&=&-\frac{m}{k}\int_0^\infty
dr\left\{U_\pi\hat{\jmath}^2_L+\left[U_\pi\left
(P^2_L-\hat{\jmath}^2_l\right)+U_{PS}P^2_l+U_SP^2_L\right]+U_CP^2_L\right\}
\nonumber \\
&\equiv&\delta_L)_{\pi L}+\left[\delta_L)_{\pi I}+\delta_L)_{PS}+\delta_L)_{S}
\right]+\delta_L)_C\;.
\label{Eq.5}
\end{eqnarray}
\noindent
In this expression, the first term represents the perturbative long range OPEP
($\pi$L), the second the iterated OPEP ($\pi$I), the third the part due to
the box and crossed-box diagrams (PS), the fourth the contribution from chiral
symmetry (S). The last one is due to the core and vanishes for waves with
$L\neq0$. In Fig.~\ref{Fig.3} we show the partial contributions to the $^1D_2$
and $^1F_3$ phase shifts as functions of energy. There it is possible to see
large cancellations of the medium range contributions, represented
by the terms within square brackets in Eq.~\ref{Eq.5}. In the case of the
$^1F_3$ wave one notes a contribution from the iterated OPEP, the
cancellation is almost complete and the total phase shift is very close to
that due to the long-OPEP term. The same patterns also holds for the $^1G_4$
and $^1H_5$ waves.

These results show that chiral symmetry is responsible for large cancellations
in the two-pion exchange interaction. This process is therefore similar to
threshold pion-nucleon or pion-deuteron \cite{RW78} scattering amplitudes,
where the main role of the symmetry is to set the scale to the problem to be
small.

\acknowledgements
M.R.R. would like to acknowledge the kind hospitality of the Division de
Physique Th\'eorique of the Institut de Physique Nucl\'eaire in Orsay, where
part of this work was performed. The work of C.A.R. was supported by FAPESP,
Brazilian Agency.

\begin{figure}
\caption{Profile functions for the bubble ($(\!)$) and triangle ($\nabla$)
scalar-isoscalar potentials and for their  chiral sum
(S), showing a strong cancellation between these two contributions.}
\label{Fig.1}
\end{figure}

\begin{figure}
\caption{Profile functions for the chiral sum (S - same as in
Fig.~\protect\ref{Fig.1}) and pseudoscalar (PS) scalar-isoscalar potentials
and for their sum, representing the full medium ranged potential (M). One sees
another strong cancellation between these two contributions.}
\label{Fig.2}
\end{figure}

\begin{figure}
\caption{Contributions for the long-OPEP ($\pi$L), iterated OPEP ($\pi$I),
pseudoscalar (PS) and chiral (S) terms of the potential to the phase shifts
for $^1D_2$ and $^1F_3$ waves. The total phase shifts are indicated by (T).}
\label{Fig.3}
\end{figure}

\end{document}